\documentclass{aa}
\usepackage{graphicx}

\usepackage{rotating}
\usepackage{amssymb,amsmath}
\usepackage{latexsym}
\usepackage{amsbsy}
\usepackage{natbib}
\usepackage{txfonts}

\bibpunct{(}{)}{;}{a}{}{,} 
\newcommand{\simless}{\mathbin{\lower 3pt\hbox
      {$\rlap{\raise 5pt\hbox{$\char'074$}}\mathchar"7218$}}} 
\newcommand{\simgreat}{\mathbin{\lower 3pt\hbox
     {$\rlap{\raise 5pt\hbox{$\char'076$}}\mathchar"7218$}}} 

\begin{document}

\title{The near-infrared reflected spectrum of source~I in Orion-KL
\thanks{
Based on observations collected at the European Southern Observatory, Chile.
  Program 076.C-0660.}
}

\author{Leonardo Testi$^{1,2}$, Jonathan C. Tan$^3$, Francesco Palla$^2$}
\institute{
    ESO, Karl Schwarzschild str. 2, D-85748 Garching, Germany
\and
    INAF--Osservatorio Astrofisico di Arcetri, Largo E.Fermi 5,
    I-50125 Firenze, Italy 
\and
    Departments of Astronomy \& Physics, University of Florida, Gainesville, FL 32611, USA
}

\offprints{ltesti@eso.org}
\date{Received ...; accepted ...}

\authorrunning{Testi, Tan \&\ Palla}
\titlerunning{The NIR spectrum of source~I in Orion-KL}

\abstract 
{Source~I in the Orion-KL nebula is believed to be the nearest example
of a massive star still in the main accretion phase.
It is thus one of the best cases for studying the properties of
massive protostars to constrain high-mass star formation theories.
Near-infrared radiation from source~I escapes through the cavity
opened by the OMC1 outflow and is scattered by dust towards our line
of sight.}  
{The reflected spectrum offers a unique possibility of observing the
emission from the innermost regions of the system and probing the
nature of source~I and its immediate surroundings.}  
{We obtained moderately high spectral-resolution
($\lambda/\Delta\lambda\sim 9000$) observations of the near infrared
diffuse emission in several locations around source~I/Orion-KL. We
observed a widespread rich absorption line spectrum that we compare
with cool stellar photospheres and protostellar accretion disk
models.}
{
The spectrum is broadly similar to strongly veiled, cool, low-gravity
stellar photospheres in the range T$_{\rm eff}\sim 3500-4500$~K,
luminosity class I-III. An exact match explaining all features has not
been found, and a plausible explanation is that a range of different
temperatures contribute to the observed absorption spectrum.
The 1D velocity dispersions implied by the absorption spectra, $\sigma\sim
30\:{\rm km\:s^{-1}}$, can be explained by the emission from a disk
around a massive, $m_*\sim 10 M_\odot$, protostar that is accreting at
a high rate, $\rm\dot{m_*}\sim 3\times 10^{-3}\,\,M_\odot {\rm yr^{-1}}$.}  
{Our observations suggest that the near-infrared reflection
  spectrum observed in the Orion-KL region is produced close to
  source~I and scattered to our line of sight in the OMC1 outflow
  cavity. The spectrum allows us to exclude source~I being a very
  large, massive protostar rotating at breakup speed.  We suggest that the
  absorption spectrum is produced in a disk surrounding a
  $\sim$10~M$_\odot$ protostar, accreting from its disk at a
  high rate of a few $\times 10^{-3}$~M$_\odot$/yr).}


\maketitle


\section {Introduction}

Our understanding of massive star formation is still very limited
compared to that of solar-mass stars, for which a generally accepted
theoretical framework exists \citep{1987ARA&A..25...23S}. The roles of
dynamical interactions and coalescence versus a more standard core and disk
accretion framework are still being debated 
\citep{2007prpl.conf..165B,2007ARA&A..45..481Z}. Progress has been limited in part by the
greater observational difficulties of studying massive protostars that
are typically more distant, embedded, crowded, and quickly evolving
than their lower mass cousins. Orion is the closest star-forming
region where young, massive (M$\ge 8$~M$_\odot$) stars are present,
and as such is an ideal test case to study observationally and to
understand theoretically.

In particular, the Kleinman-Low (KL) nebula and the Becklin-Neugebauer (BN) object
have received a lot of attention since their discovery as infrared sources 
\citep{1967ApJ...147..799B,1967ApJ...149L...1K}. 
\citet{1973ApJ...186L...7R} resolved several point sources in the KL nebula, and 
it was soon recognized that at least some of these objects were young
massive stars. Subsequent higher angular resolution, infrared observations showed
that only a limited number of these sources were self-luminous objects. 
Infrared polarimetric observations of the region have shown that one source
is responsible for the illumination of most of the nebula, with the only exception
being a small area around BN \citep{1991MNRAS.248..715M}.
Radio continuum observations have identified three sources that are likely young
(possibly still forming) massive stars in the region: source n (also associated with
an infrared source), BN, and source~I, located near the Orion hot molecular core
and near the infrared source IRc2 \citep{1995ApJ...445L.157M}.
The mid-IR spectrum of source n and its relatively low extinction imply that it is not particularly luminous \citep{1998ApJ...509..283G}. Similarly, the polarization data of \citet{1991MNRAS.248..715M} and \citet{Werner:1983fk} indicate BN is not responsible for most of the luminosity of the KL nebula. BN is known to be a runaway star \citep{1995ApJ...455L.189P}, moving through the ONC at about 30~km/s. In one scenario, it
was ejected from near the Trapezium region by dynamical interaction with the $\theta^1C$ binary
\citep{2008arXiv0807.3771T,2004ApJ...607L..47T}, and thus would have made a close passage to source I, perhaps enhancing its accretion rate by tidal interaction.
In an alternative scenario, BN was ejected by dynamical interaction with source I itself 
\citep{2005AJ....129.2281B,2008ApJ...685..333G}. Source n may also have been involved in this interaction. 
This scenario requires source I to be composed of a massive, compact binary, with total mass more than 
twice the mass of BN, i.e. $\gtrsim 20\:M_\odot$.

In either case, source I is thus expected to be the main illuminating source of the infrared nebula. 
It is associated with an outflow detected in SiO and
compact radio emission believed to originate in either a small
ionized disk
\citep{2009ApJ...698.1165G,1990ApJ...348L..65P,2009ApJ...704L..25P} or in
the ionized base of an outflow cavity \citep{2003astro.ph..9139T}.
The source is deeply embedded and is not revealed at near infrared and
shorter wavelengths. Nevertheless, the surrounding reflection nebula
offers a unique opportunity to observe the reflected emission from the
heart of the system.  \citet{1998Natur.393..340M} were the first to
obtain a low spectral resolution, near infrared (2.05-2.35~$\mu$m) spectrum of the
reflected light from source~I. They revealed that the reflected
spectrum shows the absorption lines typical of a cool photosphere,
including metal lines and the CO(2--0) overtone absorption starting at
2.29$\mu$m. The inferred equivalent photospheric temperature (T$_\star\sim 4500$~K) and
derived luminosity (L$_\star\sim 10^5$~L$_\odot$) could be consistent with either a very large,
cool, massive protostar or the ``photosphere'' of a very active
accretion disk. \citet{2000ApJ...534..976N} discus the first
possibility, using one-zone protostellar models and concluding that
the required size of $\gtrsim 300 R_\odot$ could not be
achieved by reasonable accretion rates, even up to
$10^{-2}\:M_\odot\:{\rm yr^{-1}}$. They suggest that the large size could
be attained if the protostar was rotating at near the breakup
speed, which would then imply very broadened line profiles. Subsequent
work involving multi-zone protostellar models by
\citet{2008ASPC..387..189Y} and \citet{2009ApJ...691..823H} has shown
that the sizes can be about a factor of several larger than those
predicted by the one-zone models during the (relatively brief)
luminosity wave expansion phase \citep{1986ApJ...302..590S} of the
evolution. \citet{2009ApJ...691..823H} conclude that the low temperature
implied by the infrared observations of the reflected spectrum from
source~I in Orion-KL could be achieved if the accretion rate was
$\gtrsim 4\times 10^{-3}\:M_\odot\:{\rm yr^{-1}}$, {\it averaged over the
  entire formation time of the protostar}.

However, such average accretion rates are at least an order of
magnitude greater than those expected if massive stars form from
massive cores in near equilibrium with the pressures in the Orion
Nebula region \citep{2003ApJ...585..850M}.  Since a massive protostar
will be surrounded by a relatively cool (compared to the protostar),
optically thick accretion disk, we think that this is a more likely
source of the observed near IR reflection nebula. The various models
can be tested by resolving the profiles of the absorption lines. With
this in mind we carried out intermediate-resolution (R$\sim 8900$), near-infrared
observations of the reflected spectrum from source~I in Orion-KL.

\section{Observations and data reduction}

\begin{figure*}[ht!]
\centerline{\includegraphics[width=18cm]{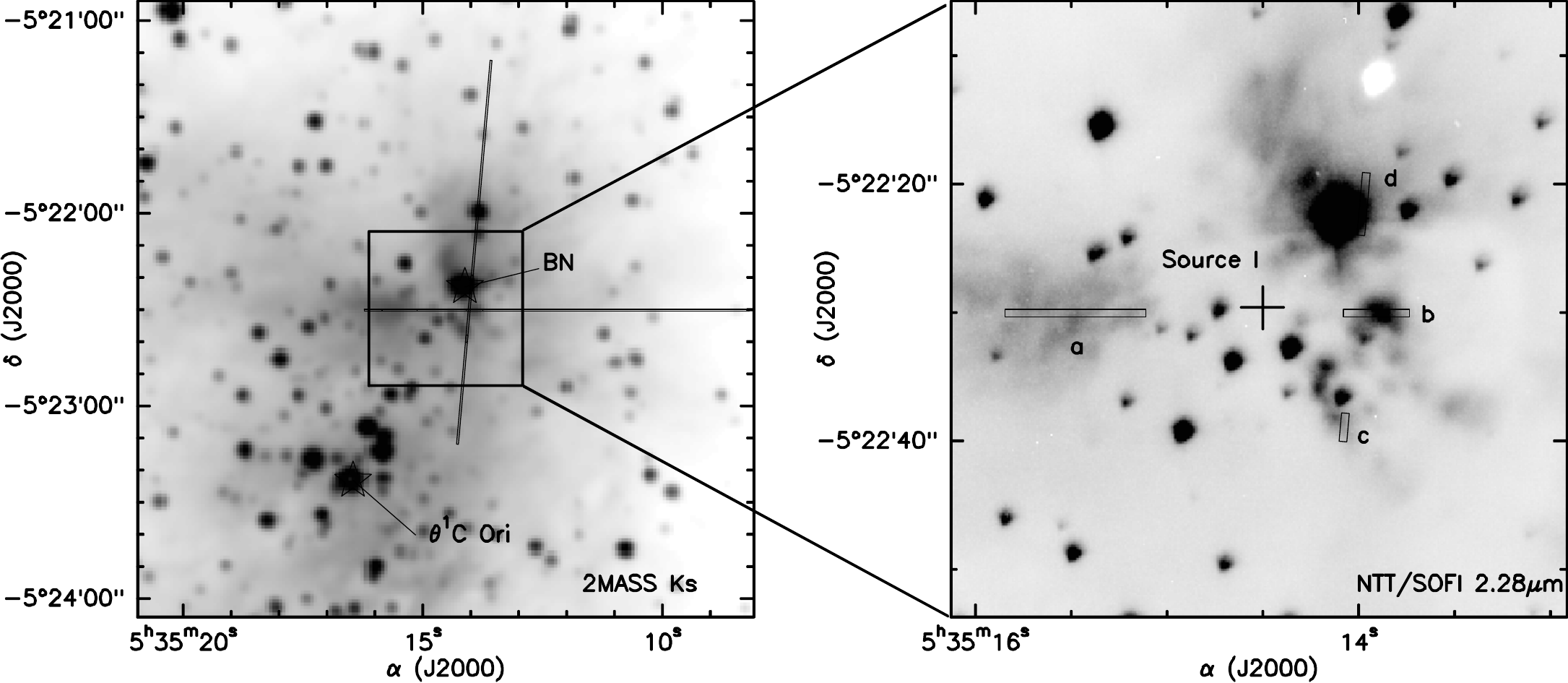} }
	\caption{Left panel: 2MASS Ks-band image of the Orion Trapezium and the
Kleinmann-Low nebula region, $\theta^1$~C and the Becklin-Neugebauer object are
marked for reference. The positions of the two on-source slits are shown. Right
panel: the region around Orion-KL source~I (marked with a cross and labeled) at
2.28$\mu$m is reproduced from an ESO-Archive SOFI-NTT observation (originally
acquired for the ESO program 64.I-0493). The small slitlets marked from a to
d show the regions we used to extract the spectra shown in Fig.~\ref{fspec}}
\label{fchart}
\end{figure*}

Near-infrared medium resolution spectra of the scattered light around the
Orion-KL region were obtained using the ISAAC near-infrared camera and
spectrograph at the ESO-VLT UT1 telescope. The observations  were carried out
in service mode on several occasions during October-November 2005. We used the
0.3 arcsec slit, and the medium-resolution grism that offered a
resolution of  $\sim$8900 across the wavelength range 2.23-2.33~$\mu$m. The observing sequence
was composed of sets of on-source and nearby (10 arcmin away) blank sky spectra
arranged as standard ABBA cycles, to allow for efficient and accurate sky
subtraction. For each of the two slit positions (see below and
Fig.~\ref{fchart}), we integrated a total of 40~min on source. Standard
calibrations (flats and lamps) and telluric standards spectra were obtained for
each observation as part of the ISAAC calibration plan. Wavelength calibration
was performed using the lamp observations and refined using the bright OH sky
lines.

The spectra were reduced using standard procedures in IRAF.  After checking
that the data sets obtained in different nights were not significantly
different, we combined all the exposures to obtain a final spectrum for each of
the two slit positions shown in the left panel of Fig.~\ref{fchart}.

\section{Results}

\begin{figure*}
\centerline{\includegraphics[width=18cm]{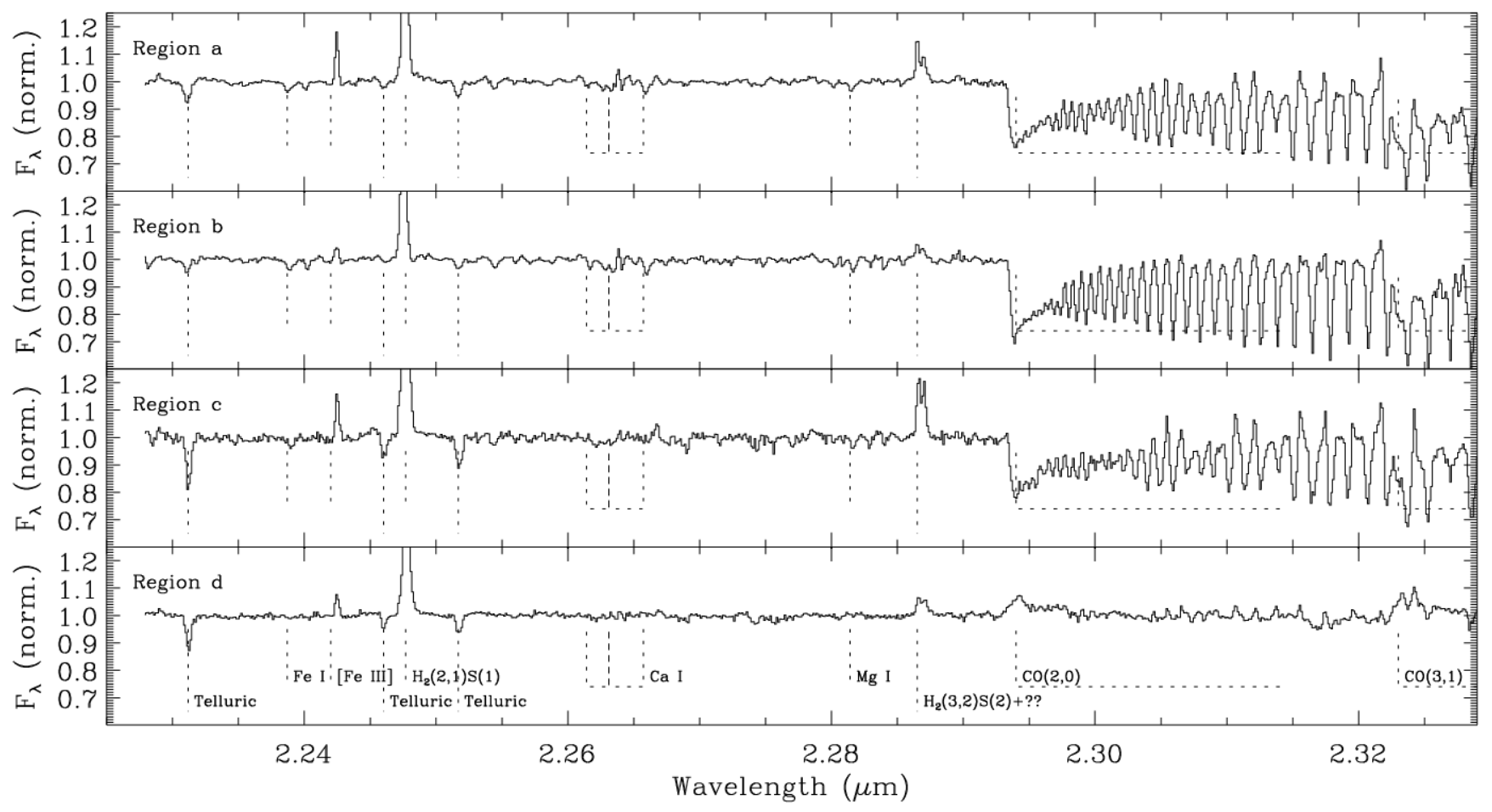} }
\caption{From top to bottom: final spectra of the four regions a to d
         shown in Fig.~\ref{fchart}. Each spectrum has been normalized to
         the continuum level. Some of the most prominent emission or
         absorption features are labeled. The features labeled ``Telluric"
         are negative residuals from the subtraction of the OH airglow
         spectrum. The spectra also show the prominent emission features of 
         H$_2$ and [FeIII] which originate from diffuse gas in the
         region and (most likely) not associated with the reflection
         nebulosity.}
\label{fspec}
\end{figure*}

Spectra were extracted for four regions corresponding to the brightest
parts of the reflection nebulosity within the observed slits (see
Fig.~\ref{fchart} for a detailed chart). Region ``a'' corresponds to
the region observed by \citet{1998Natur.393..340M}; regions ``b'' and
``c'' correspond to two bright nebulosity patches to the west and to
the south of the source~I position, respectively; and region ``d'' is very
close to the BN object.

All spectra show emission lines of H$_2$(2,1)S(1), H$_2$(3,2)S(2)
(blended with an unidentified infrared emission feature), and
[FeIII]. These emission lines are most likely associated with the
Orion HII region \citep{1998A&A...330..696M,2000A&A...364..301W}
and/or the OMC1 molecular hydrogen outflow 
\citep[e.g.][]{1996ApJ...467..676C,2000PASJ...52....1K,2007A&A...466..949N}.
Some contamination from the molecular hydrogen emission can 
also be seen superposed on the CO overtone absorption.
For the analysis of the reflected absorption spectrum, we
have subtracted the nebular spectrum. We estimated the average 
nebular spectrum by averaging regions of the long slits
away from the reflected emission and then subtracted it
from each of the absorption spectra so as to remove the
H$_2$(3,2)S(2) line completely.

The spectrum of region ``d'' is very different from the other three because
it hardly shows any absorption feature, and the CO lines are detected
in emission.  This is consistent with the infrared spectra of the BN
object \citep{1979ApJ...232L.121S}, and confirms that the nebular
emission close to the BN object is dominated by the reflected light
from this source rather than source~I within the central KL region, as
also inferred from the polarization maps of this region
\citep{1991MNRAS.248..715M}.  Therefore, we do not consider this spectrum
further in the analysis.

The polarization images of regions ``a'', ``b'', and ``c'' suggest that
the emission here is dominated by scattered light from the embedded
source or sources within the central KL region
\citep{1991MNRAS.248..715M}. The three spectra (Fig.~\ref{fspec}, top
panels) show a common absorption line spectrum. In Fig.~\ref{fspec} we
have marked the identification of the most prominent absorption
lines. In particular, along with the CO(2,0) and CO(3,1) ro-vibrational
bands, we clearly detect the CaI triplet at $\sim 2.265\,\mu$m, the
MgI$\lambda 2.281\,\mu$m, and the FeI$\lambda 2.239\,\mu$m lines.

As mentioned above, our spectra from regions ``a'', ``b'', and ``c''
show a virtually identical absorption line pattern. This result
confirms and strengthens the conclusion by \citet{1998Natur.393..340M}
that the absorption spectrum does not originate in the diffuse
interstellar medium along the line of sight, but that we are instead
observing the absorption spectrum of a single source that is scattered
towards us by dust along different lines of sight towards the
KL region.

\subsection{Comparison with cool stellar photospheres}

\citet{1998Natur.393..340M} points out that the reflected absorption
spectrum they observed from the region roughly overlapping with our
spectrum ``a'' has an absorption line pattern similar to that of a
cool stellar photosphere. Their analysis suggests that the absorption
spectrum, especially the atomic lines, is consistent with that of the
giant star HD12014 veiled by an excess featureless continuum
emission, which contributes about half of the continuum emission at
$\sim$2~$\mu$m. HD12014 has a K0Ib photosphere, corresponding to an
effective temperature of $\sim$4500~K and low gravity.

\begin{figure*}
\centerline{\includegraphics[width=14cm]{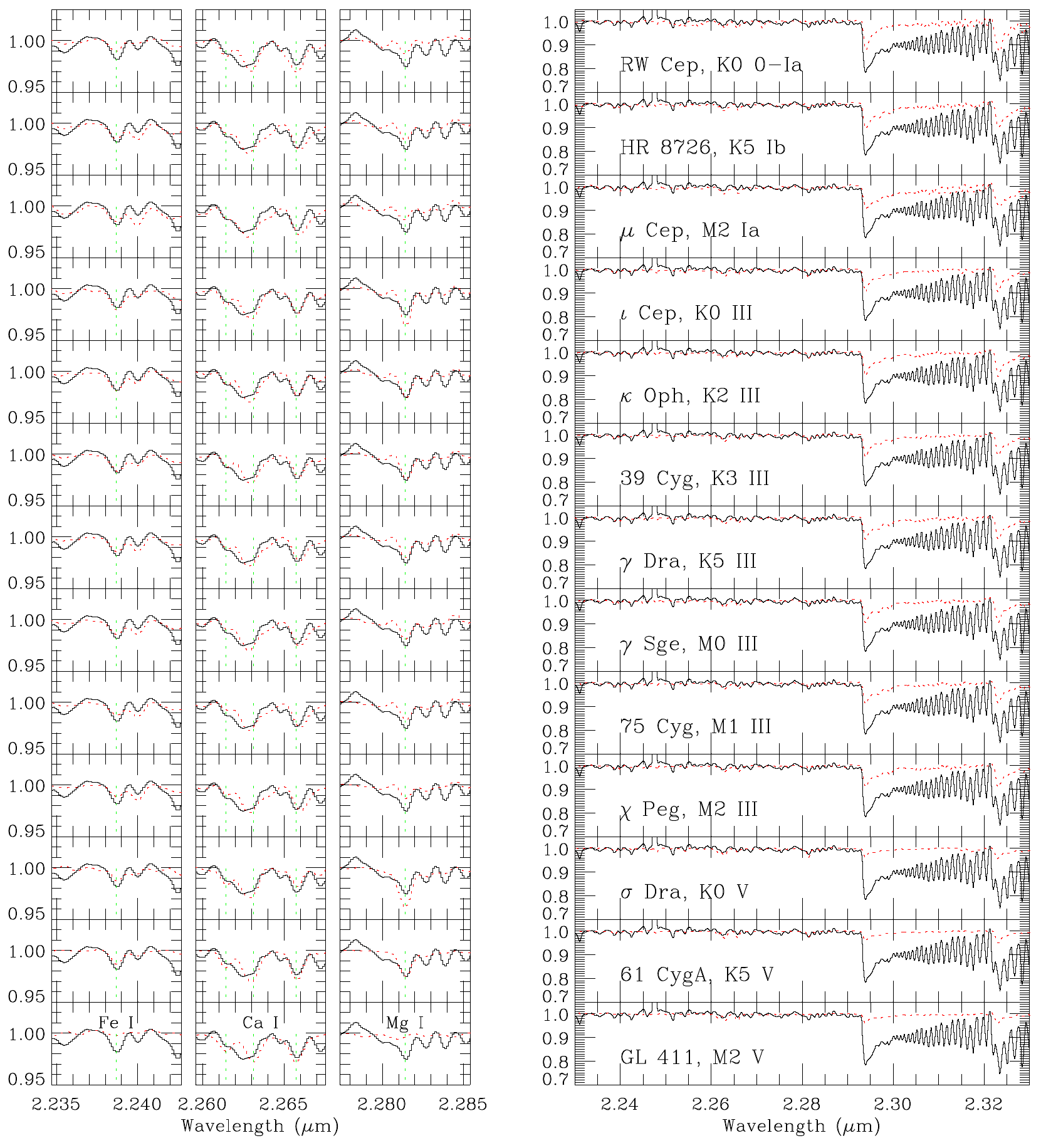} }
	\caption{Comparison between our average reflected spectrum (solid line) and the 
	stellar spectra from \citet[red dotted lines]{1986ApJS...62..501K}. In the left panels we show 
	enlargement around the
	lines of Fe~I, Ca~I, and Mg~I (the wavelength of each feature is marked with a vertical green dotted line), 
	and on the right side we show the full wavelength range of our spectrum, and the 
	name and spectral type of the stars are marked in the right side panels. 
	}
\label{fkh86}
\end{figure*}

Our higher resolution spectra allow a more detailed comparison with 
stellar photosphere spectral libraries.
%
%
%
In Fig.~\ref{fkh86} we show the comparison between the spectra in the
library of \citet{1986ApJS...62..501K} and our average reflected
spectrum, obtained by combining the data from regions ``a'', ``b'', and
``c''. 
Our spectrum was smoothed to the same resolution as the spectra in the
library. The photospheric spectra 
were veiled in such a way to approximately match the depth of the central line of the Ca~I triplet 
in the Orion spectrum. To match the Ca~I absorption the amount of veiling required by the late type
spectra is such that the stellar photosphere contributes $\sim 20\%$ of the computed spectrum, while for the 
earlier spectral types the photosphere dominates the spectrum.

The ratio of the atomic absorption lines is best matched with veiled
stellar photospheres of spectral types M2-K0, corresponding to
effective temperatures in the range 3500-4500~K.  Later types have much more pronouced CaI
absorption than the MgI, while the opposite is
true for earlier spectral types.  
In all cases, as noted by
\citet{1998Natur.393..340M}, the CO absorption bands are much more
pronounced in the Orion spectrum than in the stellar
photospheres. In this respect, the lowest gravity, later type spectra
in the library are closer to our observed spectrum.  This suggests
that the spectrum may be consistent with an even lower gravity
``photosphere'', such as from an accretion disk.

\subsection{Velocity dispersion of the absorption spectra}

\begin{figure*}
\centerline{\includegraphics[width=16cm]{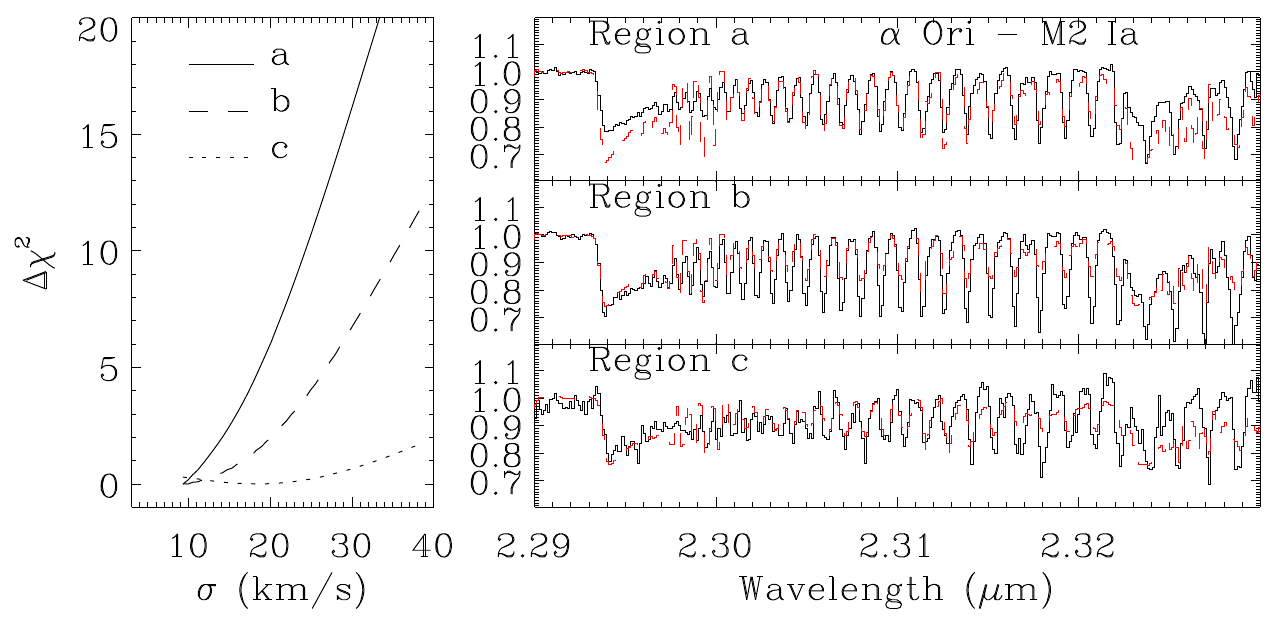} }
	\caption{Results of the $\chi^2$ fits to the CO absorption in Orion with the high-resolution 
	infrared spectrum of $\alpha$~Ori \citep{WH96}. In the left side panel we show the $\Delta\chi^2$
	curves as a function of  the $\Delta\rm V$
	used to smooth the library spectrum. On the right we show the comparison between the observed 
	spectrum in each of the three regions (full lines) with the best-fitting photosphere (dashed lines, corresponding 
	to the $\Delta\chi^2$ minima in the left-hand side plot).}
\label{fchi2_wh96}
\end{figure*}

To estimate the broadening of the absorption line spectra, we compared
the CO absorption in our ISAAC spectra with the CO absorption in the
high-resolution spectra of \citet{WH96}. The spectra from the library
were convolved with a Gaussian of variable width to mimic line
broadening (in addition to that intrinsic to the stellar 
library), shifted in velocity, veiled and then smoothed at the same
resolution and resampled onto the same wavelength grid of our ISAAC
spectra for proper comparison. These parameters were allowed to vary
and a $\chi^2$ cube was constructed. In Fig.~\ref{fchi2_wh96} we show,
for the fit to the spectrum of $\alpha$~Ori, the variation of the
$\chi^2$ as a function of the 1D velocity dispersion ($\sigma=\rm
FWHM/1.665$)
of the Gaussian smoothing function for each of the three regions
``a'', ``b'', and ``c''. The results obtained using this star are
consistent with those of the other late type, low-gravity stars in the
library (such as $\alpha$~Her or RX~Boo).  The results show that to
match our spectra, the photospheric templates need to be smoothed with
Gaussian profiles with $\sigma\le 30$~km/s for region ``c'', while
the spectra of regions ``a'' and ``b'' require a much smaller
smoothing, $\sigma\le 15$~km/s. 
The true, total velocity
dispersions will be greater than these limits owing to the contribution
of the intrinsic dispersion in the stellar atmospheres used in the
spectral library.

Another estimate of the line broadening can be done directly with a
simple modeling of the CO lines. The line parameters were taken from
the HITRAN database \citep{2009JQSRT.110..533R}, and line intensities
were computed following \citet{1996PhDT........68H} using a simple LTE
approximation. As in the procedure with the template spectra, we
convolved the model spectra with a Gaussian profile of a given full
width at half maximum, and the synthetic spectrum was then scaled to match
the observed absorption spectrum, smoothed at the appropriate spectral
resolution, and resampled on the same wavelength grid as our ISAAC
spectra. This simplified analysis does not allow us to accurately reproduce
all the CO line intensities, but the line broadening
required to match the observed spectra is consistent with the results
derived using the spectral templates. 
In particular region ``a'' requires $\sigma \lesssim 20$~km/s, while
``b'' and ``c'' are consistent with $\sigma \sim 27 \pm 3$~km/s.  

We repeated the analysis using a double-peaked line profile as
expected from a disk \citep[e.g.][]{1988ApJ...325..231K}. Although we do not
observe double-peaked lines, our observations would still be
consistent with line profiles where the double peaks
are separated by $\le 45$~km/s.

The estimates of the linewidths are also consistent with recent observations of
mid infrared CO absorption lines toward a region close to our 
region ``b'' \citep[][their source IRC3]{2010arXiv1001.0650B}.
Their spectrum also shows emission components,
most likely from outflowing gas, which further complicates the analysis of the
line profile.

\section{Discussion}

Our higher resolution spectra (R$\sim$8900 vs. $\sim$1000) allow us to
test the different scenarios put forward by
\citet{1998Natur.393..340M}, \citet{2000ApJ...534..976N}, and 
\citet{2009ApJ...691..823H}  on the nature of the source producing the absorption
spectrum. The latest authors have attempted to explain the observed,
cool spectrum as being produced by a very large ($r_*\gtrsim
200\:R_\odot$), hence cool ($T<5500$~K), protostar, which requires
very high accretion rates averaged over the whole formation time ($\dot{m}_*\gtrsim 4\times
10^{-3}\:M_\odot\:{\rm yr^{-1}}$), and the protostar to be at a
particular stage in its evolution ($m_*\simeq 20\:M_\odot$),
undergoing swelling due to the ``luminosity wave'' \citep{1986ApJ...302..590S}
immediately before the onset of Kelvin-Helmholz contraction.

For a protostar forming from a near-equilibrium gas core, such high
average accretion rates would require very high ambient pressures,
equivalent to the self-gravitating weight of a gas clump with
$\Sigma\gtrsim 44\:{\rm g \:cm^{-2}}$
\citep{2003ApJ...585..850M}. However, as described by these authors,
the Orion Nebula Cluster has a much lower mean mass surface density of
$0.24\:{\rm g \:cm^{-2}}$. While moderate enhancements in $\Sigma$,
pressure, and thus $\dot{m}_*$ are expected in the central regions of
the ONC, it appears to be difficult to create the conditions that
would allow such high accretion rates to produce the observed NIR
scattered light directly from the protostellar surface.
The instantaneous accretion rate from a turbulent core is expected to
show significant fluctuations about the mean value. Fluctuations may
also be induced by tidal interactions with passing stars in the
cluster. This scenario is discussed below in the context of source I
and the BN object. However, these fluctuations are not expected to
significantly change the size of the star, which is set by the
accretion rate averaged over the last growth time ($m_*/\dot{m}_*$) of
the protostar.

The large protostar scenario makes predictions for the velocity
dispersion of the CO banheads as a function of the rotational velocity
of the protostellar surface relative to its breakup velocity. For
example a 20~$M_\odot$, 200~$R_\odot$ protostar rotating at 50\% 
of the Keplerian velocity at its
surface has an equatorial rotation speed of 69~$\rm
km\:s^{-1}$. Viewed at an inclination angle of $45^\circ$, this would
produces an observed line profile with a velocity dispersion of
$\sigma_{\rm rot}\simeq 50\rm km\:s^{-1}$ \citep[e.g.][]{1992oasp.book.....G}.


While the large protostar scenario is still marginally consistent with
the observed velocity dispersion, we favor the alternative scenario of
the NIR spectrum showing CO bandhead absorption being produced in an
active accretion disk. There are several arguments in support of this:
(1) accretion disks are expected to be present around protostars; (2)
the large protostar scenario requires very high accretion rates and a very
special protostellar evolutionary phase; (3) the detection of prominent
low excitation CO absorption near 4.6~$\mu$m in the reflected spectrum
from region ``b'' (IRC3) by \citet{2010arXiv1001.0650B} suggests there
is a range of ``photospheric'' temperatures. This is expected in
an active disk, but difficult to reconcile with a single
stellar photospheric temperature --- in fact, these low excitation lines
are not prominent in cool stellar photospheres
\citep[e.g.]{2003ApJS..147..379S}; (4) there is evidence that source I is
in an active accretion phase, such as the intense outflow activity
from the region (e.g. the OMC1 outflow) and the claims that the cm
radio emission and SiO maser motions around source I indicate the
presence of either a compact circumstellar disk 
\citep{2007ApJ...664..950R,2010ApJ...708...80M} or the ionized base of
an outflow cavity (Tan \& McKee 2003).

The accretion disk scenario was also briefly discussed by \citet{2000ApJ...534..976N}.
To further explore this possibility, in the next section we compare
our spectrum with a more detailed model of a massive protostar-disk 
system as expected in the turbulent core model for high-mass star 
formation \citep{2003ApJ...585..850M}.

There are claims that the outflow and disk from source I are oriented orthogonally to the orientations assumed in our model, i.e. with the outflow aligned with a NE-SW axis
\citep[e.g.][]{2009ApJ...698.1165G,2007ApJ...664..950R}. Such an orientation is hard to reconcile with the observed orientation of the large-scale outflow from the KL region 
\citep{1993Natur.363...54A,1996ApJ...467..676C} and the morphology of the NIR nebula. 
We suspect there may still be the possibility of misinterpreting of the outflow and disk orientations from the maser data since the maser emission may not provide a uniform sampling of the gas in these structures. Alternatively, if there has been a strong, recent dynamical interaction of source I with BN
\citep{2005AJ....129.2281B,2004ApJ...607L..47T}, then the orientation of the disk may have been changed. However, it is still difficult to understand how direct illumination of the NW and SE regions of the KL 
nebula could be achieved by source I in this scenario, so for these reasons we favor the fiducial 
model as we have described it.


\subsection{Comparison with the turbulent core protostellar model}

\citet{2003ApJ...585..850M} have estimated the range of protostellar models
consistent with producing the observed luminosity of the
Orion Hot Core, i.e. the Kleinmann-Low Nebula, of $\sim 1-5\times
10^4\:L_\odot$. Here we consider a comparable range of models, with
parameters listed in Table~\ref{tab:models}. They involve a
60~$M_\odot$ gas core forming a star with 50\% efficiency and at
various stages of collapse, i.e. with $m_*$ in the range of 8 to
20~$M_\odot$. If the core was in pressure equilibrium with a $\Sigma =
1\:{\rm g\:cm^{-2}}$ surrounding clump, then the accretion rates would
be a few $\times 10^{-4}\:M_\odot\:{\rm yr^{-1}}$. 

It is possible that the Orion KL protostar has suffered a tidal
perturbation in the past $\sim 1000$~yr due to the close passage of
the runaway BN star \citep{2004ApJ...607L..47T}.  This is expected to
have enhanced the accretion rate through the inner disk and thus also
the protostellar mass outflow rate, perhaps causing the apparently
``explosive'' morpholgy of the outflow \citep{1993Natur.363...54A}.
Thus we also consider some models with accretion rates enhanced by a
factor of 10, especially for lower mass protostars that will still
satisfy the observed total luminosity constraint.

For the 15~$M_\odot$ protostar accreting at $\dot{m}_*=3.28\times
10^{-4}\:M_\odot\:{\rm yr^{-1}}$, the star has a mean spherical radius of
$7.42\:R_\odot$ and internal luminosity of $L_*=2.01\times
10^4\:L_\odot$ with a photospheric temperature of $25,400$~K 
\citep{2003ApJ...585..850M}.  The
accretion boundary layer has a luminosity of $L_{\rm BL}=1.03\times
10^4\:L_\odot$ (with a photospheric temperature of $32,000$~K). The
accretion disk is assumed to have the same luminosity as that of the
boundary layer. These estimates of accretion luminosity should be
regarded as upper limits since they do not include any reduction from
the powering of protostellar outflows, which may have mechanical
luminosities of about 50\% of the total accretion power.

  \begin{table*}
 \caption[]{Parameters of protostellar models}
 \label{tab:models}
\begin{tabular}{ccccccccc}
  \hline
  $m_*$ & $r_*$ & $L_*$ & $T_*$ & $\dot{m}_*$ & $L_{\rm BL}$ & $L_{\rm tot}$ & $\sigma$ & $\sigma_{\rm irr}$\\
 ($M_\odot$) & ($R_\odot$) & ($L_\odot$) & (K) & ($M_\odot\:{\rm yr^{-1}}$) & ($L_\odot$) & ($L_\odot$) & ($\rm km\:s^{-1}$) & ($\rm km\:s^{-1}$) \\
  \hline

8.0 & 12.0 & 5270 & 12200 & $2.40 \times 10^{-4}$ & 2450 & 10200 & 61.0 & 52.7\\
8.0 & 12.0 & 5270 & 12200 & $2.40 \times 10^{-3}$ & 24500 & 54300 & 40.4 & 36.0\\
10.0 & 12.5 & 5850 & 14400 & $2.68 \times 10^{-4}$ & 3300 & 12500 & 64.5 &  55.9\\
10.0 & 12.5 & 5850 & 14400 & $2.68 \times 10^{-3}$ & 33000 & 72000 & 42.7 &  38.0\\
15.0 & 7.42 & 20100 & 25400 & $3.28 \times 10^{-4}$ & 10300 & 40600 & 69.3 & 56.6\\
15.0 & 7.42 & 20100 & 25400 & $3.28 \times 10^{-3}$ & 103000 & 226000 & 46.5 & 39.0\\
20 & 5.81 & 44800 & 35000 & $3.78\times 10^{-4}$ & 20200 & 85200 & 73.8 & 57.3 \\
  \hline
  \end{tabular}
  \end{table*}

\begin{figure}
\centerline{\includegraphics[height=9cm]{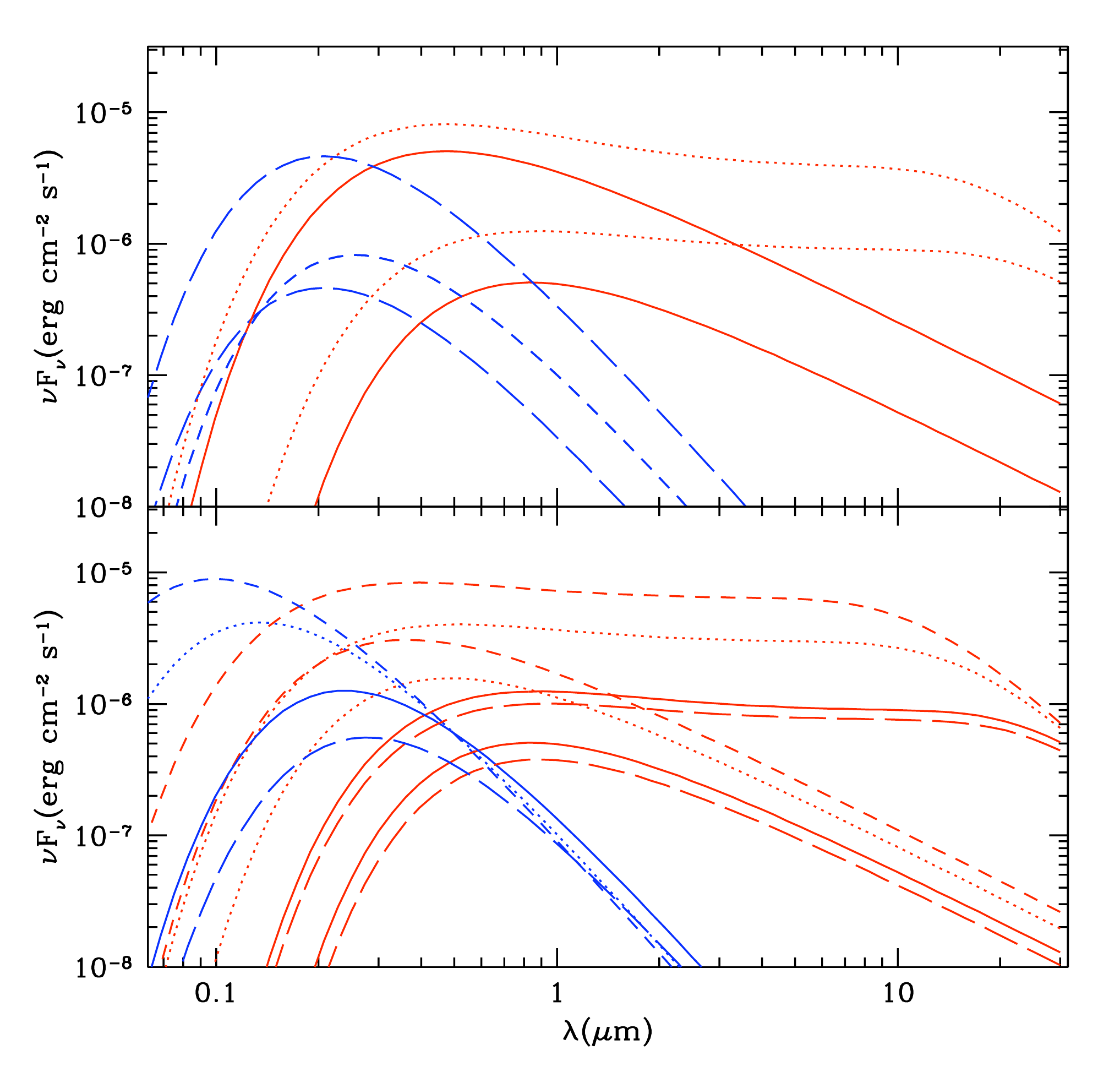} }
\caption{Spectra of protostellar models (see text). Top panel: For
 the fiducial accretion rate model with $m_*=10\:M_\odot$, the fluxes from the star and
 boundary layer are shown by the dashed and lower
 long-dashed lines, respectively. The lower solid line shows
 the flux from the accretion disk (integrating out to 5000$r_*$) assuming negligible protostellar
 heating, while the lower dotted line shows the case accounting for
 protostellar heating. The upper long-dashed, solid, and dotted lines
show the equivalent quantities for the model with the accretion rate enhanced
by a factor of 10. Bottom panel: The solid lines show for the
 fiducial $m_*=10\:M_\odot$ model the flux from the star$+$boundary
 layer, nonirradiated and irradiated accretion disks. The long-dashed, dotted,
 and dashed lines show the equivalent fluxes from the fiducial
 accretion rate models with
 $m_*=8,15, and 20\:M_\odot$, respectively. In all cases, the flux at
 2.3~$\rm \mu m$ is dominated by disk emission.}
\label{fig:diskspec}
\end{figure}

We assume the protostar has a geometrically thin, optically thick
accretion disk. We consider the case of a purely active disk,
i.e. with negligible heating of the disk by the protostar, and the
case of a protostellar-heated disk, following the treatment of \citet{2005ApJ...628..817M} 
and \citet{2006MNRAS.373.1563K}. The reprocessing factor, $f_{\rm irr}$
is defined as the ratio between the incident flux $F_{\rm irr}$ normal
to the disk surface, and the spherical stellar flux at that radius:
$F_{\rm irr}=f_{\rm irr}L/(4\pi r^2)$. An analytic approximation for $f_{\rm irr}$ is
$f_{\rm irr}\simeq 0.1 \epsilon ^{-0.35}$, where $\epsilon$ is the instantaeous
star formation efficiency from the core, which we have taken to be
50\% \citep{2000ApJ...545..364M}. In both the non-irradiated and
irradiated cases, emission from the accretion disk dominates the
emergent NIR flux, as shown in Fig.~\ref{fig:diskspec}.

\begin{figure}
\centerline{\includegraphics[height=9cm]{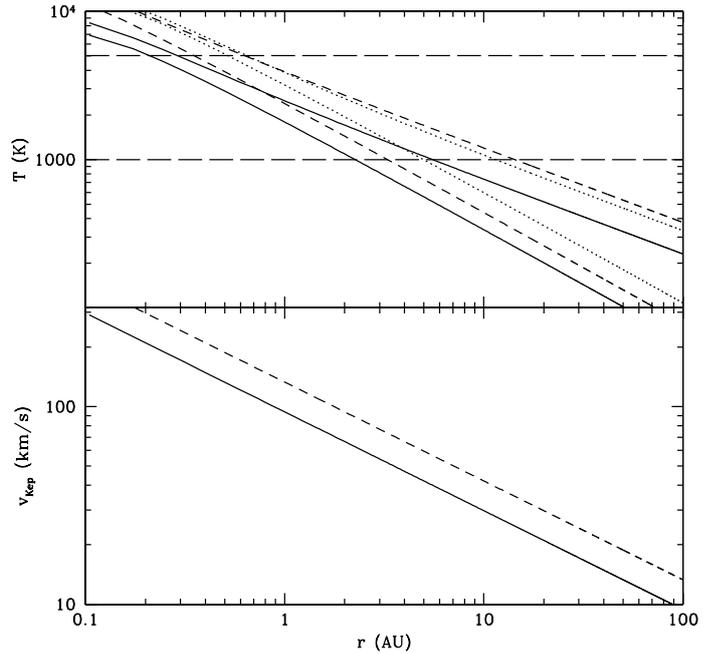} }
\caption{Radial profiles of protostellar accretion disk models. Top
 panel: Disk photospheric temperature for non-irradiated disk (lower
 solid line) and irradiated disk (upper solid line) for
 $m_*=10\:M_\odot$ with fiducial accretion rate.  The equivalent models for $m_*=10\:M_\odot$ with accretion rate enhanced by a factor of 10 are shown by the dotted lines. The equivalent models for $m_*=20\:M_\odot$ with fiducial accretion rate
 are shown by the dashed lines. The long-dashed
 horizontal lines show the approximate range of temperatures where
 we consider there are contributions to the CO bandhead
 equivalent width. Bottom panel: Keplerian velocities for the
 $m_*=10, 20\:M_\odot$ models, shown by the solid and dashed
 lines, respectively.}
\label{fig:profile}
\end{figure}

\begin{figure}
\centerline{\includegraphics[height=9cm]{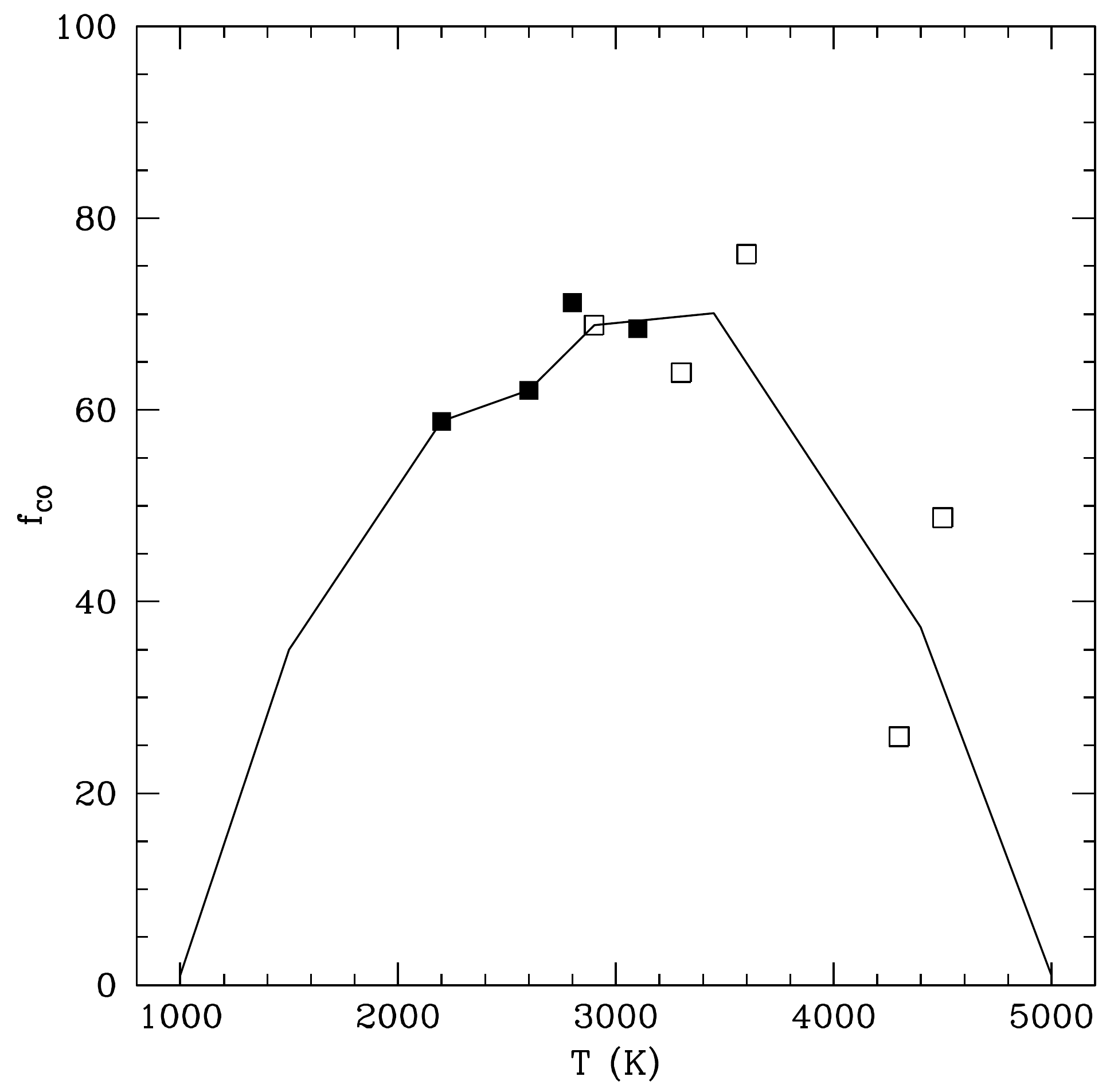} }
\caption{Weighting factor of CO(2-0) bandhead equivalent width over
  the wavelength range $2.292 - 2.322\:{\rm \mu m}$ derived from
  giant/supergiant standard spectra (open squares) and Mira spectra
  scaled by a factor of 0.7 (solid squares).}
\label{fig:fCO}
\end{figure}

The radial profiles of disk photospheric temperature and Keplerian
velocity are shown in Fig.~\ref{fig:profile}. We evaluated the
contribution to the total CO line profile from each emitting part of
the disk, weighting by the disk area, flux at $2.3\:{\rm \mu m}$, and
a temperature-dependent ``CO bandhead weighting factor'', $f_{\rm
  CO}(T)$, which is derived from the observed equivalent widths of
giant and supergiant standard stars (see Fig.~\ref{fig:fCO}). This
factor is uncertain at $T<2000$~K, although we have checked various
functional forms that decline with temperature for this temperature
range and found that our results are not particularly sensitive to
such variations. We assume a typical inclination angle of $45^\circ$
of dust particles to the rotation axis of the accretion disk, since
the P.A. of the axis of the bipolar flow (and the observed near IR
reflection nebulosity) is approximately from the NW to the SE
\citep[e.g.][]{1996ApJ...467..676C}.
The application of this outflow geometry to source I is
consistent with the interpretation of source I's radio emission as
the ionized base of an outflow cavity \citep{2003astro.ph..9139T},
but not with the interpretation that the disk axis is
oriented from the NW to SE \citep{2007ApJ...664..950R,2010ApJ...708...80M}.

The 1D velocity dispersions, $\sigma$, of the resulting line profiles
are listed in Table~\ref{tab:models}. The irradiated disks have
lower values caused by the disk being warmer at a given radius and
the CO emitting region being located at larger radii. If an
inclination angle of $30^\circ$ is adopted, for example, for the
$15\:M_\odot$ model with $\dot{m}_*=3.28\times 10^{-4}\:M_\odot\:{\rm
  yr^{-1}}$, then the velocity dispsersion drops to $\sigma = 49.1\:{\rm
  km\:s^{-1}}$ and $\sigma_{\rm irr}=40.1 \:{\rm km\:s^{-1}}$.

The models with enhanced accretion rates cause the disk to be warmer
at a given radius, thus pushing the CO bandhead region of the disk to
larger radii and lower Keplerian velocities. This produces lower
values of $\sigma$.

We conclude that the fiducial turbulent core protostellar models
(i.e. those without enhanced accretion rates) for the Orion KL
protostar predict CO bandhead 1D observed velocity dispersions that
are in the range 53 to 75~$\rm km\:s^{-1}$. The velocity dispersion is
not very sensitive to $m_*$ because the temperature range that
contributes to CO bandhead absorption, i.e. $T\sim 2000 - 4000$~K
occurs at a fairly constant range of Keplerian velocities in the
accretion disk, nearly independent of $m_*$. Disk irradiation raises
the temperature of the disk at any given radius, thus shifting the
zone that contributes to CO bandhead equivalent width to larger radii
and thus yielding narrower line profiles. The velocity dispersion of
the line profile is somewhat sensitive to the adopted inclination
angle between the scattering surfaces in the outflow and the disk
rotation axis. The fiducial protostellar models predict 1D CO bandhead
velocity dispersions that are about a factor of two to three greater
than observed.
A better match is achieved if there is a
relatively stronger contribution to CO bandhead EW from lower velocity
dispersion gas, which can be reproduced in models with accretion rates
enhanced by a factor of $10$.
Higher velocity resolution observations of the scattered light from
Orion KL are needed to better establish the significance of the
observed velocity dispersion in order to further test theoretical
models, including the search for double-peaked line profiles.

\section{Summary and conclusions}

In this paper we have presented medium-resolution (R$\sim 8900$), near-infrared
spectra of the reflection nebulosity in Orion-KL. Our main results
are as follows.

\begin{itemize}
\item 
The absorption line spectra in various regions of the reflection
nebula in Orion-KL are consistent with each other, except for a
small difference in the line broadening. This
suggests a common origin for the absorption line close to the region
emitting the continuum emission. We thus confirm the earlier
suggestion by \citet{1998Natur.393..340M} that the absorption spectrum
is generated close to the position of source~I, either from the
protostellar photosphere or the surrounding accretion disk.
This also suggests that the outflow cavity from the main illuminating
source in the KL nebula, i.e. source I, is aligned from the SE to the
NW, which is difficult to explain in models that invoke a disk plane
along this axis \citep{2007ApJ...664..950R,2010ApJ...708...80M}.

\item 
The observed absorption spectra are broadly consistent with
low-gravity, late-type, heavily veiled stellar photospheres with
spectral types in the range M2-K0 and luminosity class I-III,
corresponding to effective temperatures in the range
3500-4500~K. Nevertheless, it was not possible to find a good match
for all the observed absorption features with a single photospheric
template. In particular, the observed CO absorption in the Orion
spectra are much deeper than in the stellar photospheres that match
the atomic lines. Although it is possible that an even lower gravity
photosphere than those used in the comparison may alleviate or solve
this inconsistency, it is likely that this discrepancy could be solved
if the Orion spectrum is produced by the combination of a wide range
of photospheric temperatures, as in a very active accretion disk.

\item 
The observed line profiles are consistent with being unresolved at our
spectral resolution in regions ``a'' and ``b'' and marginally resolved
in region ``c''. These results imply $\sigma\sim 30$~km/s for region
``c'' and about a factor of two less for regions ``a'' and ``b''.
We thus exclude that the spectra could be originated from the
photosphere of a massive ($\sim 20 M_\odot$), large ($\sim 200
R_\odot$) protostar rotating close to the breakup speed.  The
linewidths are consistent with the predictions of a model of an active
accretion disk around a massive protostar (M$\sim 10$~M$_\odot$). 
Models with accretion rates
enhanced above the fiducial values are favored in order to reproduce
the relatively narrow line widths.
\end{itemize}

We conclude that the most likely source of the near infrared
absorption spectrum is the accretion disk surrounding the massive
protostar source~I, with the system in a state of relatively high
accretion rate.

\acknowledgements{We thank the ESO support astronomers and La Silla Paranal Observatory staff
for their support in the preparation and execution of the service mode observations
for this program. This work was partially supported through ASI grants to the
INAF-Osservatorio Astrofisico di Arcetri. JCT acknowledges support from NSF CAREER grant
AST-0645412 and the ESO visitor program.}

\bibliographystyle{aa}
\bibliography{ltesti}













\end{document}